\documentclass[reprint,superscriptaddress,amsmath,amssymb,aps,prb,floatfix]{revtex4-1}
\usepackage{graphicx}
\usepackage{dcolumn}
\usepackage{bm}
\usepackage{upgreek}
\usepackage{lineno}
\usepackage[normalem]{ulem}
\usepackage{color}

\begin{document}

\title{GPU-accelerated Monte Carlo simulations of anisotropic Heisenberg ferromagnets }

\author{Michalis Charilaou}
\email{michalis.charilaou@louisiana.edu}
\affiliation{Department of Physics, University of Louisiana at Lafayette, Lafayette, Louisiana 70504}

\date{\today}

\begin{abstract}
The Monte Carlo method is a powerful technique for computing thermodynamic magnetic states of otherwise unsolvable spin Hamiltonians, but the method becomes computationally prohibitive with increasing number of spins and the simulation of real materials and nanostructures is cumbersome. This paper presents the acceleration of Monte Carlo simulations of the three-dimensional anisotropic Heisenberg model on Graphics-Processing Units (GPU). The GPU implementation of the method presented here provides an acceleration of two orders of magnitude over conventional implementations and enables the simulation of large systems, with any crystal lattice, containing up to $10^8$ spins on a single GPU. This offers the possibility to simulate complex structures and devices that are hundreds of nanometers in size in order to compute their magnetic state at finite temperature with atomistic resolution.
\end{abstract}
\maketitle

The Heisenberg model \cite{heisenberg1928} is one of the most successful and insightful ways to describe ferromagnetism and magnetic phase transitions. The model considers each atom as a dipole and assigns an interaction between neighbors, such that the internal energy of the system is minimized when the dipoles are aligned. In the quantum limit, the dipoles are treated as quantum spin operators, but with increasing number of eigenstates the density of states becomes increasingly continuous, and the model can be treated classically, where each spin is a classical vector \cite{skomski2020}. The Heisenberg model has been used to study phase transitions in ideal 3D vector systems \cite{domb1962,ritchie1972,ferer1986,nightingale1988,landau1991}, but it can also serve as a basis to model real magnetic materials and nanostructures. In insulators, the model for localized spins can be applied in a straight-forward way and offers a realistic description of the physics. In conductors, particularly metals where electrons are delocalized, the model can also be applied, but the intrinsic material parameters need to be adjusted by reducing the effective exchange due to the interatomic electron hopping and having non-integer spin moments due to the fact that electrons are shared between atoms. Hence, the classical Heisenberg model is an excellent approximation for both insulators and conductors, i.e., oxides and metals, and it enables the modeling of magnetic states in materials with atomistic resolution. 

The Hamiltonian of the anisotropic Heisenberg model contains contributions from the ferromagnetic exchange interaction $J$ and the anisotropy energy density $K$
\begin{equation}
\mathcal{H}=\!-J\sum_{ij}^N \mathbf{S}_i\cdot \mathbf{S}_j - K\sum_i^N \left(\mathbf{S}_i\cdot\mathbf{\hat{e}}\right)^2 
\end{equation}
with the three-dimensional spin vector $\mathbf{S}=\left(S_x,S_y,S_z\right)$ and the unit vector $\mathbf{\hat{e}}$ of the anisotropy easy axis, for the case of uniaxial anisotropy. 

Analytical solutions for the Heisenberg model are possible only in one or two dimensions \cite{mermin1966,joyce1967} or by employing mean-field approximations. Hence, numerical solutions are required to model systems containing more than a few spins. 

The Monte Carlo method \cite{binder1986,landaubinder} is a powerful technique for numerically finding the equilibrium state of a system and it has become an established method for solving spin models \cite{janke1993,nowak2000,vargas2002,kechrakos2003,vargas2006,milde2013,buhrandt2013,charilaou2014,charilaou2015,chen2017}. With the Monte Carlo method, the spin system is placed in a heat bath and the order parameter is computed at each temperature. The single-spin update Metropolis algorithm \cite{metropolis1953} finds the thermal equilibrium by repeating these steps: (i) a spin is chosen at random; (ii) a new orientation is proposed randomly; (iii) the change of energy $\Delta \mathcal{H}$ is calculated and if $\Delta\mathcal{H}\leq 0$ the change is accepted, whereas if $\Delta\mathcal{H}> 0$ the change is accepted if $P\leq e^{-\beta\Delta\mathcal{H}}$, where $\beta$ is the inverse temperature and $P$ is a random number between 0 and 1. Typically, to achieve thermal equilibrium, $10^4$ Monte Carlo steps per spin (MCS) are required. Once the system reaches thermal equilibrium, arithmetic averages of the spin vectors of accepted states are taken and the magnetization is 
\begin{equation}
\mathbf{M}=\frac{1}{N\left|S\right|}\left<\sum_i^N\mathbf{S}_i\right>\; ,
\end{equation}
with $M_\mathrm{s}=N\left|S\right|$ the saturation magnetization.

As with any computational method, the technique becomes increasingly slower with increasing system size. However, advances in computing now enable new implementations of the technique with substantial acceleration. Specifically, the use of Graphics-Processing Units (GPU) is extremely promising for accelerating simulations and for the development of machine learning and neural networks \cite{CudaBook2012}. The potential of using GPU for Monte Carlo simulations has been discussed in several works, both for the Ising model \cite{preis2009,block2010,weigel2011,weigel2012} and the 2D Heisenberg model \cite{weigel2012,weigel2018}, as well as the 3D Ising \cite{parisi2015} and 3D Heisenberg spin glass \cite{parisi2011}. Optimization of GPU memory usage has been proposed by lattice decomposition, such as the checkerboard decomposition of 2D lattices \cite{weigel2011,weigel2012}, but that type of lattice decomposition limits the applicability of the technique to simple square lattices. For the simulation of real magnetic materials, however, it is important to be able to simulate three-dimensional systems with lattices with higher symmetry, i.e., beyond the simple square or simple cubic lattice.

Here, I present an implementation of the Monte Carlo method for the anisotropic Heisenberg model with GPU acceleration, enabled by Nvidia's CUDA framework \cite{cuda}. In contrast to previous GPU-accelerated implementations, this method enables the simulation of spin systems with any crystal structure and has a speed-up of two orders of magnitude compared to the conventional CPU implementation. 

The code was written in C++ and makes use of the internal cuRAND library, which uses a Mersene-Twister-type pseudorandom number generator. The basic principle of this implementation is that the entire spin system is linearized into three one-dimensional arrays, one for each component of the spin vector, which reside on the GPU's global memory, and each GPU block handles a part of that array. At each step, $N_b$ number of blocks are called to perform a spin update in parallel. Each block handles $n$ number of spins, and within each block one of those spins is selected at random to be updated at every step. The energy difference between current state and proposed state is computed locally on each block, and subsequently the spin vectors are updated on the global GPU memory. Once the system is in thermal equilibrium, averages of the spin state are taken every $N$ steps using parallel reduction \cite{cook2013}, and once the simulation is complete the magnetization average is copied from the GPU to the CPU to be exported. An alternative option would be to make use of shared memory instead of relying on global memory, because on-chip shared memory of each block is faster, but it has been shown that there is little advantage of doing that \cite{parisi2011}.

The number of thread blocks is limited by the hardware, i.e., the type of GPU, and by the number of spins in the system. If the number of spins handled by each block is too small, the calculation of the energy, which is performed locally on each block, will be concurrent with the update of spins involved in that calculations by other blocks, and this will lead to the loss of continuity. In this investigation, the optimal number of spins per block was found to be 8, i.e. a system with $N=128^3$ (or $2^{21}$) spins will be simulated on $2^{18}$ blocks. Additionally, parallelization of Monte Carlo is possible when we only consider short-range interactions. Long-range dipole-dipole interactions cannot be implemented, because after each Monte Carlo step $N/n$ spins have been updated independently and the local dipolar field acting on each spin changes drastically at each step, and the Markov chain is broken. However, long-range interactions can be implemented in the form of effective shape anisotropy \cite{guimaraes2009}.

The simulations presented in this paper were performed with a Tesla V100 (32 GB) GPU, access to which was provided by the Louisiana Optical Network Infrastructure. Additional tests were performed with a Titan Xp (12 GB) and a GeForce MX150 (2GB) GPU on a laptop computer. The spin system was mapped on to a simple cubic lattice, but as noted above it can be done on any lattice symmetry because of the linearized spin arrays. In order to test the performance of the implementation, simulations were performed for different number of thread blocks $N_b$, in the range $2^{11}-2^{15}$, and different lattice sizes, in the range $N=16^3-512^3$. For the thermalization of the system, $10^4$ MCS were performed and additional $10^4$ MCS were processed to acquire thermal averages of the magnetization. For the purpose of generality, the exchange energy was fixed at $J=1$, the anisotropy at $K=0.01$, and the spin magnitude at $\left|S\right|=1$.

\begin{figure}
\includegraphics[width=1.0\columnwidth]{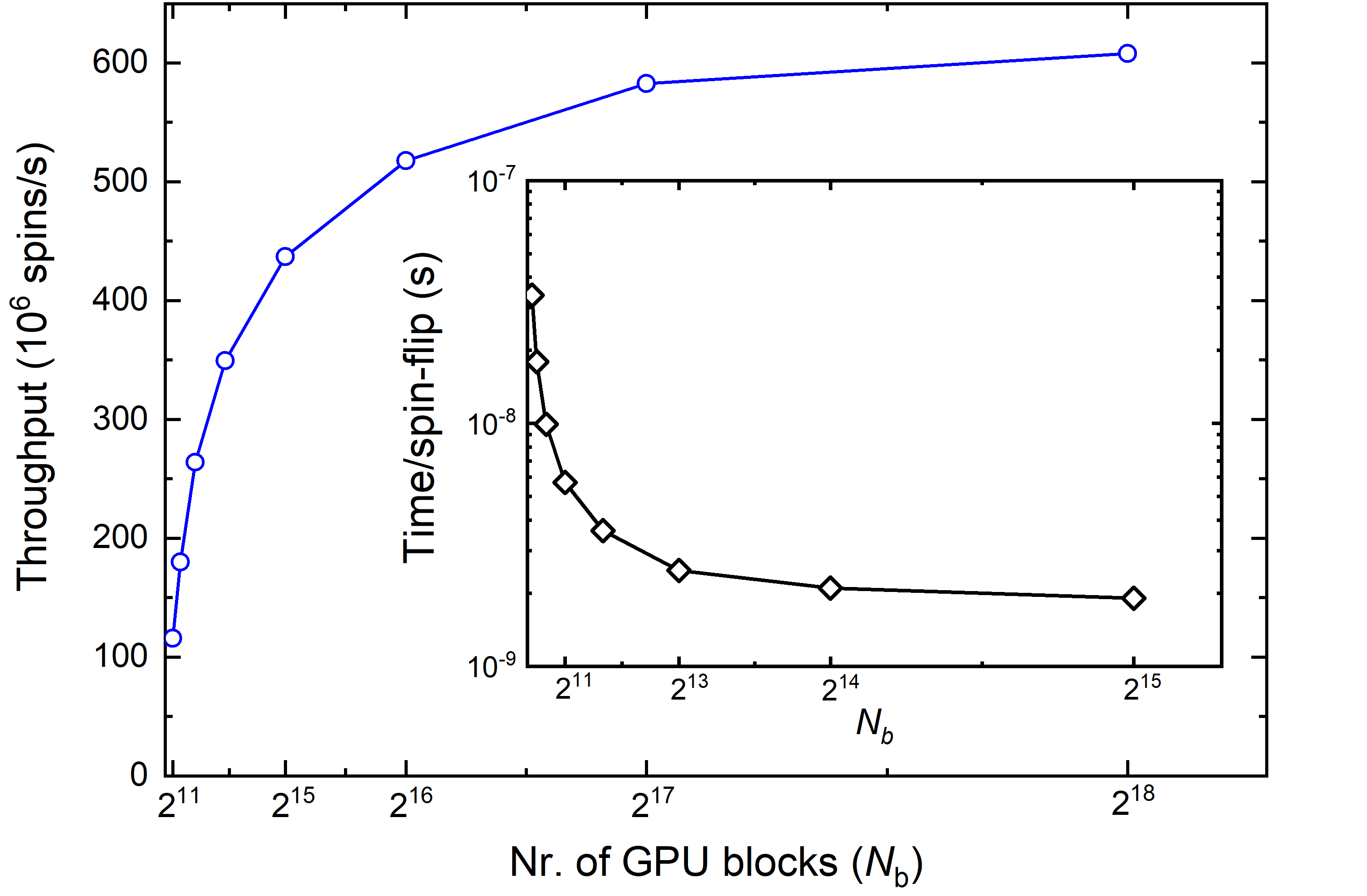}
\caption{Performance enhancement as a function of parallelization: the simulation throughput for a system with $N=128^3$ increases with the number of GPU blocks, whereas (inset) the time per spin flip decreases by nearly two orders of magnitude. } 
\label{fig:parallel}
\end{figure}

Figure \ref{fig:parallel} shows the performance of the simulation as a function of $N_b$ for a system with $N=128^3$ spins at a temperature of $T=0.9T_\mathrm{C}$, where $T_\mathrm{C}$ is the Curie temperature. With increasing number of thread blocks, and decreasing number of spins handled by each thread block ($n$), the throughput of the simulation increases rapidly for $N_b=2^{10} - 2^{15}$, and for $N_b\geq 2^{16}$ it approaches saturation. This trend is associated with the efficiency of the random number generator and the bandwidth of data transfer and depends on the GPU architecture. The maximum throughput was obtained for $N_b=2^{18}$ at $610\times 10^6$ spin updates per second, which corresponds to 1.63 ns per spin update, and it is two orders of magnitude faster than conventional CPU implementations, which require a time on the order of 100 ns per spin \cite{weigel2012}. Notably, even with the smallest GPU, a peak throughput of $40\times 10^6$ spins per second was achieved (not shown here), which is an order of magnitude faster than the conventional CPU implementation, therefore these GPU-accelerated simulations can be performed even on laptop computers with a CUDA-compatible GPU without the need to access high-end devices or large-scale facilities. 

\begin{figure}
\includegraphics[width=1.0\columnwidth]{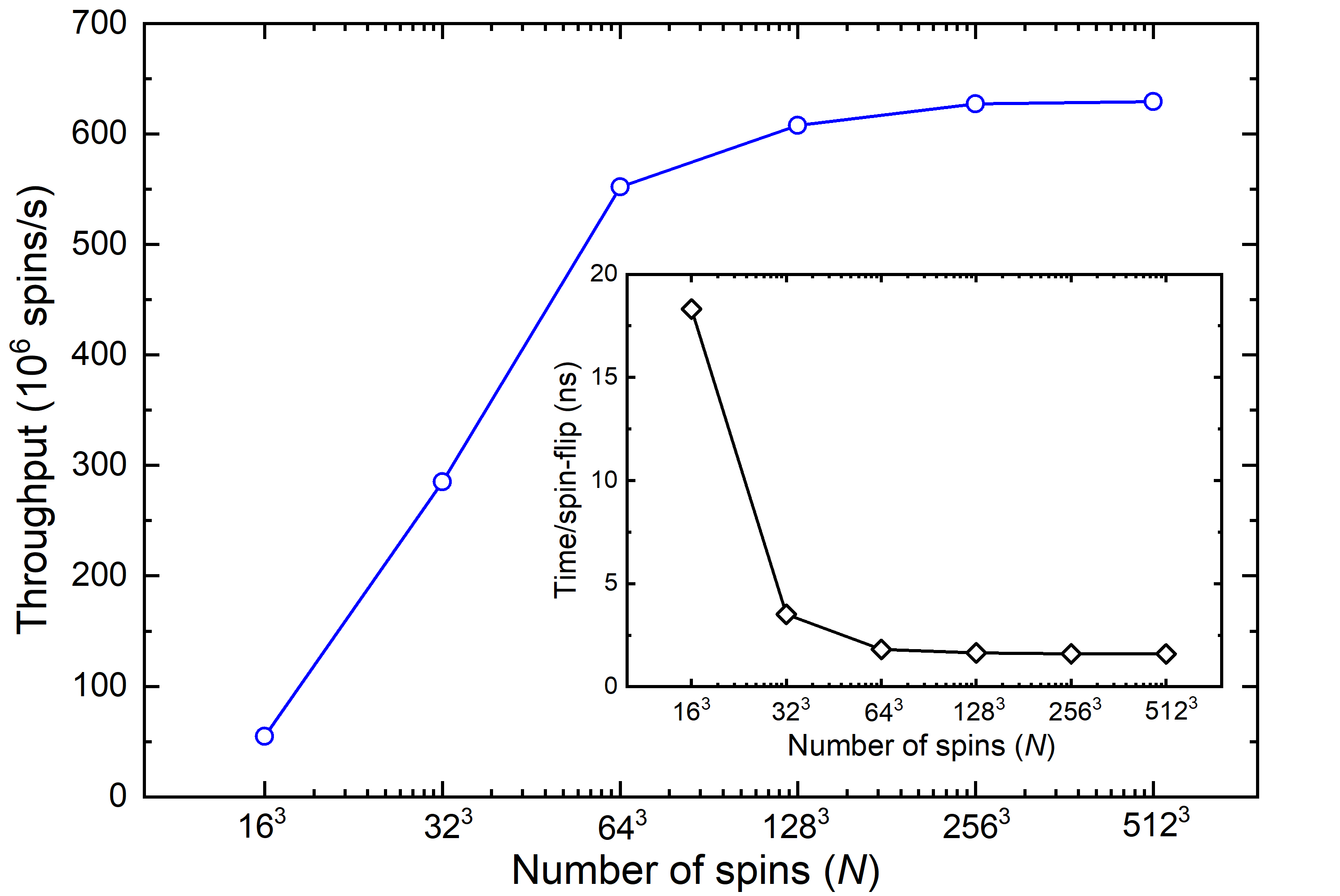}
\caption{Acceleration of the Monte Carlo simulation expressed as throughput (in $10^6$ spin flips per second) as a function of lattice size. The inset shows the time per spin-flip, which decreases with increasing system size with a minimum of 1.55 ns per spin flip.} 
\label{fig:acceleration}
\end{figure}
The acceleration of the computation also depends on the total number of spins in the simulation system ($N$). With increasing $N$ the use of the thread blocks becomes increasingly efficient and for systems with $10^3 - 10^6$ spins the throughput increases rapidly and approaches saturation, reaching $630\times 10^6$ spin updates per second for a system with $512^3$ spins, as shown in Fig. \ref{fig:acceleration}. This throughput corresponds to a time of 1.59 ns per spin update. The upper limit of the system size is set by the size of the spin arrays, which depends on the hardware and operating system and is typically at 2 GB. For a system with $512^3$ spins, each float array occupies a size of 0.54 GB, whereas for a system with $1024^3$ spins each array would occupy 4.3 GB of memory. These limitations can be circumvented by partitioning the lattice in different arrays and sequentially thermalizing one array while storing the rest on dynamic random access memory (DRAM) or the hard-drive, but this would slow down the simulation due to the delay while transferring data from GPU to CPU and back. Despite this, the system with a size of $512^3$ corresponds to a nanocube with a side length on the order of 100 nm for typical ferromagnetic structures, and a simulation of a structure this size with atomistic resolution has not been reported, to the extent of my knowledge. \\

\begin{figure}[h!]
\includegraphics[width=0.99\columnwidth]{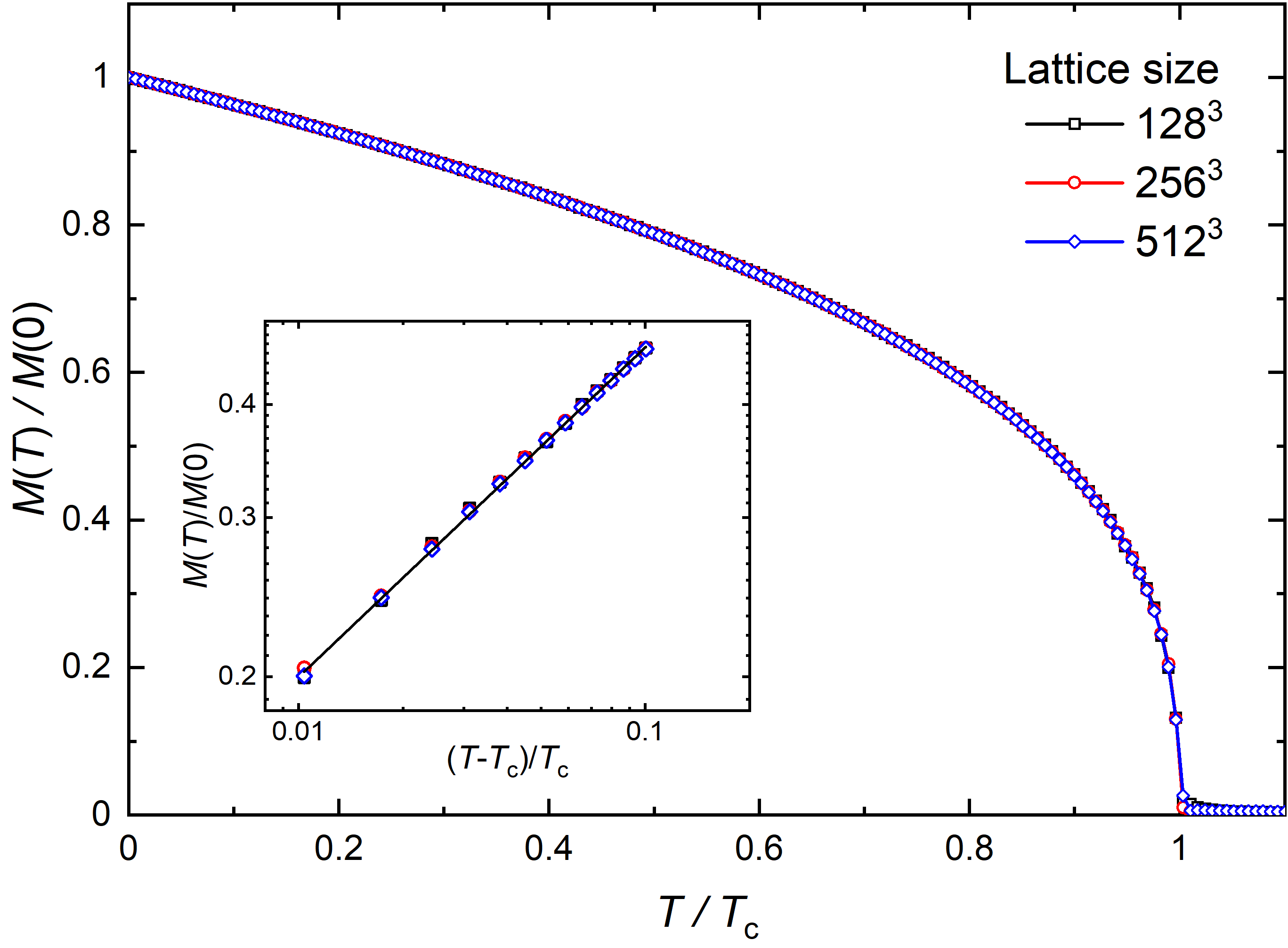}
\caption{Simulated magnetization as a function of temperature for lattice sizes of $128^3$, $256^3$, and $512^3$, showing numerical stability of the method for systems containing up to $10^8$ spins. The inset shows the magnetization as a function of reduced temperature $t=(T-T_\mathrm{c})/T_\mathrm{c}$ near the critical region in a log-log plot, where the linear fit yields a critical exponent $\beta=0.364\pm 0.002$.} 
\label{fig:thermo}
\end{figure}

In order to test the accuracy of the implementation, simulations of the magnetization as a function of temperature were performed for different system sizes, as shown in Fig. \ref{fig:thermo}, and from these simulations the critical temperature and the critical exponent were extracted. Figure \ref{fig:thermo} shows the characteristic $M(T)$ curve of the Heisenberg model on the simple cubic lattice with periodic boundary conditions. The critical temperature is $T_\mathrm{C}/J=1.445$ and the critical exponent of the reduced temperature $t=(T-T_\mathrm{C})/T_\mathrm{C}$ in the vicinity of the phase transition, where the magnetization follows the power law $M=t^\beta$ (see inset to Fig. \ref{fig:thermo}), is $\beta=0.364\pm 0.002$, in excellent agreement with high resolution Monte Carlo simulations ($T_\mathrm{C}/J=1.445$ and $\beta=0.362$) \cite{landau1991,landau1993} and field theory ($\beta=0.3645$) \cite{guillou1977,guillou1980}. Hence, the implementation presented here demonstrates not only a drastic acceleration but also a high numerical stability for systems containing up to $10^8$ spins.  

In conclusion, the presented GPU implementation of the Monte Carlo method for the anisotropic Heisenberg model, free from limitations on the crystal structure, enables rapid and large-scale simulations of magnetic materials. Simulations can be performed on any CUDA-compatible GPU device, and the capacity to model systems containing up to $10^8$ spins allows for the simulation of entire nanostructures or devices with atomistic resolution at finite temperature, which is particularly important for the development of novel spintronic devices for non-volatile data storage.

The author gratefully acknowledges funding from the Louisiana Board of Regents [contract Nr. LEQSF(2020-23)-RD-A-32] and is thankful to Leonardo Pierobon for fruitful discussions. Portions of this research were conducted with high performance computational resources provided by the Louisiana Optical Network Infrastructure (http://www.loni.org).

%

\end{document}